\documentclass[preprint]{aastex631} 

\usepackage{makecell}
\usepackage{booktabs}
\usepackage{hyperref}
\usepackage{nameref}
\usepackage{lineno}
\usepackage{url}   
\usepackage{appendix}  

\hypersetup{linkcolor=red,citecolor=cyan,filecolor=blue,urlcolor=magenta}
\usepackage{subfigure}
\usepackage{color}
\shorttitle{JVAS J1311+1658: A New $\gamma$-Ray–Emitting Compact Symmetric Object}
\shortauthors{Jiang et al.}

\begin{document}

\title{First Detection of $\gamma$-Ray Emission from the Compact Symmetric Object JVAS J1311+1658}

\correspondingauthor{Da-Ming Wei}
\email{ dmwei@pmo.ac.cn}

\author[0009-0004-3221-2603]{Xiong Jiang}
\affiliation{Key Laboratory of Dark Matter and Space Astronomy, Purple Mountain Observatory,
Chinese Academy of Sciences, Nanjing 210023, People’s Republic of China}
\affiliation{School of Astronomy and Space Science, University of Science and Technology of China, Hefei, Anhui 230026, People’s Republic of China}

\author[0009-0002-4513-4486]{Yang-Ji Li}
\affiliation{Yunnan Observatories, Chinese Academy of Sciences, Kunming 650216, People’s Republic of China}
\affiliation{University of Chinese Academy of Sciences,  Beijing 100049, People’s Republic of China}

\author[0009-0005-1848-0553]{Hai Lei}
\affiliation{School of Physics and Materials Science, Guangzhou University, Guangzhou 510006, People’s Republic of China}

\author[0000-0002-9758-5476]{Da-Ming Wei}
\affiliation{Key Laboratory of Dark Matter and Space Astronomy, Purple Mountain Observatory,
Chinese Academy of Sciences, Nanjing 210023, People’s Republic of China}
\affiliation{School of Astronomy and Space Science, University of Science and Technology of China, Hefei, Anhui 230026, People’s Republic of China}


\begin{abstract}
We report the first detection of $\gamma$-ray emission from the young radio galaxy JVAS~J1311+1658, classified as a compact symmetric object (CSO). This detection is characterized by a recent GeV $\gamma$-ray flare identified in Fermi-LAT data during MJD~60032.6--60132.6, with a $\gamma$-ray source detected at a significance level of $\sim6.2\sigma$. The average 0.1--300~GeV flux is measured to be $(1.6 \pm 0.6)\times10^{-8}\,\mathrm{ph\,cm^{-2}\,s^{-1}}$, with a photon spectral index of $\Gamma = 2.15 \pm 0.185$. We find that a radiative model of the radio lobes significantly underestimates the observed $\gamma$-ray emission. The strong flux and short-term variability over $\sim$100 days suggest that the emission likely originates from newly launched sub-kiloparsec-scale jets at the core. This detection provides a unique window into the extreme environments and early-stage jet activity of young radio galaxies, offering insights into their initial evolution and the formation of relativistic jets in the earliest phases of galaxy growth.
\end{abstract}

\keywords{Galaxy jets; Active galactic nuclei; Gamma-ray sources}

\section{Introduction} \label{sec:intro}

The presence of $\gamma$-ray emission in galaxies points to the existence of extreme physical conditions. Blazars—active galactic nuclei (AGNs) with relativistic jets nearly aligned with our line of sight—dominate the extragalactic $\gamma$-ray sky due to strong Doppler boosting of their high-energy radiation \citep{2015A&ARv..24....2M, 2023arXiv230712546B}. The latest incremental release of the Fourth Catalog of AGNs detected by Fermi Large Area Telescope (Fermi-LAT) (4LAC-DR3; \cite{2022ApJS..263...24A}) includes 3,405 AGNs at Galactic latitudes $|b| > 10^\circ$, with the vast majority ($\sim$ 98$\%$) classified as blazars and only $\sim$ 2$\%$ as radio galaxies. Radio galaxies are considered to be the misaligned counterparts of jetted active AGNs, with larger jet inclination angles ($> 10^\circ$) than blazars. Due to the large viewing angle, the radiation emitted by a radio galaxy is less boosted compared to that observed from blazars. With their misaligned jets, radio galaxies provide a unique opportunity to investigate non-thermal processes occurring in the unbeamed regions of AGNs, which are typically dominated by beamed emission from the jet in blazars. 

CSOs are a distinct class of radio galaxies characterized by their symmetric, sub-kiloparsec-scale radio structures \citep{2016AN....337....9O, 2024ApJ...961..242R, 2024ApJ...961..241K}. They are considered important because they are thought to be the progenitors of large-scale radio galaxies, whose lobes can extend over tens to hundreds of kiloparsecs \citep{1996ApJ...460..612R}. These intriguing sources are believed to represent an early stage in the evolutionary sequence of radio galaxies, with kinematic ages typically estimated to be less than a few thousand years \citep{1996ApJ...460..612R, 2016AN....337....9O}. Their compact sizes may result from their youth, a dense galactic environment that inhibits jet propagation, and/or episodic or recurrent nuclear jet activity \citep{2021A&ARv..29....3O}. There have been theoretical predictions suggesting that CSOs could be relatively strong $\gamma$-ray emitters, as their compact radio lobes contain abundant highly relativistic particles. The $\gamma$-rays may be produced either via the interaction of these relativistic electrons with low-energy optical–UV photons from the accretion disk, or through hadronic processes \citep{2008ApJ...680..911S, 2011MNRAS.412L..20K}. In addition, theoretical studies also suggest that during the initial phase of expansion, the lobes have electron temperatures of roughly 1 GeV, which can generate GeV $\gamma$-ray emission via bremsstrahlung processes \citep{2009MNRAS.395L..43K}.

To date, only a few CSOs have been detected in $\gamma$ rays, mainly because their jets are typically misaligned with respect to the line of sight, resulting in small Doppler factors and weak high-energy emission \citep{2021MNRAS.507.4564P}, and because of the limited sensitivity of early Fermi-LAT observations. \citet{2014ApJ...780..165M} initially reported no $\gamma$-ray counterparts at the locations of 12 young radio sources, but later reported a Fermi-LAT detection near the position of NGC 6328 ($z = 0.0144$; \citealt{2014MNRAS.440..696A}, associated with 4FGL J1724.2-6501; \citealt{2016ApJ...821L..31M}), making NGC 6328 the first CSO known to emit $\gamma$-ray radiation. Subsequent studies, aided by the steadily increasing sensitivity of the Fermi-LAT with accumulated exposure and improved event reconstruction, have gradually identified $\gamma$-ray emission from several other CSOs, including TXS~0128+554 ($z = 0.0365$; \citealt{2012ApJS..199...26H}, associated with 4FGL~J0131.2+5547; \citealt{2020ApJ...899..141L}), NGC~3894 ($z = 0.0107$; \citealt{2002MNRAS.329..877C}, the counterpart of 4FGL~J1149.0+5924; \citealt{2020A&A...635A.185P}), 4C~+39.23B ($z = 0.121$; \citealt{1998MNRAS.295..946R, 2022ApJ...927..221G}), and DA~362 ($z = 0.260$ \footnote{It should be noted that although \citet{1992PASA...10..140W} reported a redshift for DA~362, the origin of this value remains unclear. Moreover, \citet{2024ApJ...961..240K} conducted a systematic investigation of the spectroscopic redshifts for this source, but in their study, DA~362 was listed as having no known redshift. Therefore, the redshift value of 0.260 should be treated with caution.}; \citealt{1992PASA...10..140W}, associated with 4FGL~J1416.0+3443; \citealt{2025ApJ...979...97S}).
 Among them, NGC 6328, TXS 0128+554, and NGC 3894 have not shown significant flux variability so far \citep{2023ApJS..265...31A}, which is consistent with theoretical predictions: radio lobes typically expand at sub-relativistic velocities, and thus the $\gamma-$ray emission is not expected to exhibit significant flux variations, especially on short timescales (i.e., weeks to months). However, flaring activity was observed in the monthly binned $\gamma$-ray light curve of DA 362, indicating that the $\gamma$-ray emission likely originates from the core or inner jet rather than from the radio lobes \citep{2025ApJ...979...97S}. The differences in $\gamma-$-ray behavior suggest that the $\gamma-$ray emission environment of CSOs is still not well understood. Therefore, it is essential to expand the sample size of these peculiar sources, as this is crucial for studying the radiative mechanisms in their jets, examining their interactions with the surrounding environment, comparing them with CSOs that lack $\gamma$-ray detection, and gaining a deeper understanding of their evolutionary processes.

In order to obtain a genuine sample of CSOs for astrophysical studies, \cite{2024ApJ...961..240K} recently compiled a comprehensive catalog of 79 bona fide CSOs by reviewing the literature and analyzing their multifrequency radio observations. In order to identify potential $\gamma$-ray emitting CSOs, and motivated by the flaring $\gamma$-ray activity observed in DA 362, we conducted a systematic analysis of monthly binned Fermi-LAT data for the CSOs among the 79 bona fide sources that had not previously exhibited $\gamma$-ray emission. As a result, we discovered a new GeV $\gamma$-ray source at the position of JVAS J1311+1658 (also known as ICRF J131123.8+165844), making it only the sixth $\gamma$-ray-detected object among the class of bona fide CSOs.

In this Letter, we report the first-ever detection of $\gamma-$ray emission from the nearby CSO radio galaxy JVAS J1311+1658 \citep{2016MNRAS.459..820T, 2024ApJ...961..240K} using Fermi-LAT. We adopt a $\Lambda$CDM cosmology with $ \Omega_{M} $ = 0.32, $ \Omega_{\Lambda} $ = 0.68, and a Hubble constant of $H_{0}$ = 67 km$^{-1}$ s$^{-1}$ Mpc$^{-1}$ \citep{2014A&A...571A..16P}.

\section{Data Analysis and Results} \label{sec:2}
\subsection{Fermi-LAT Data} \label{subsec:1}
The Fermi-LAT is a $\gamma$-ray instrument designed to detect photons through their conversion into electron–positron pairs. It operates over a broad energy range, from 20 MeV to beyond 300 GeV. The LAT consists of three main components: a high-resolution converter tracker that determines the direction of incoming $\gamma$-ray, a CsI(Tl) crystal calorimeter that measures their energy, and an anticoincidence detector that helps discriminate against charged particle background \citep{2009ApJ...697.1071A}.

For this analysis, we examined 16.4 years of Pass 8 Fermi-LAT data, spanning from 2008 August 4 to 2025 January 1, within the energy range of 0.1–300 GeV ({\tt evclass} = 128, {\tt evtype} = 3; \citealt{2018arXiv181011394B}). The data reduction and analysis were carried out using {\tt Fermitools}\footnote{\url{https://fermi.gsfc.nasa.gov/ssc/data/analysis/documentation/}} version 2.0.8 and the accompanying {\tt Fermitools-data} v0.18.
A circular region of interest (ROI) with a $10^\circ$ radius centered on the radio coordinates of CSOs was adopted. To minimize contamination from Earth-limb photons, a zenith angle cut of $90^\circ$ was applied, along with standard quality filters ({\tt DATA\_QUAL==1 \&\& LAT\_CONFIG==1}). 
The parameters of background sources within the ROI, as well as the normalizations of the Galactic diffuse ({\tt gll\_iem\_v07}) and isotropic ({\tt iso\_P8R3\_SOURCE\_V3\_v01}) templates\footnote{Available at \url{https://fermi.gsfc.nasa.gov/ssc/data/access/lat/BackgroundModels.html}}, were left free in the fits. Other source parameters were fixed to values listed in the 4FGL-DR4 catalog\footnote{\url{https://fermi.gsfc.nasa.gov/ssc/data/access/lat/14yr_catalog/}}; \cite{2023arXiv230712546B}).
To search for potential excess $\gamma$-ray emission near the positions of CSOs, we placed a “test source” at the radio coordinates of each CSO and assessed its significance by comparing the likelihoods of two models: the original optimized model and the model with the additional test source. Denoting $L_0$ as the likelihood of the original model and  $L$ as the likelihood of the model including the test source, the test statistic (TS) for the test source is calculated as follows:TS = $-2\ln(L_0/L)$ \citep{1996ApJ...461..396M}. If the test source is significantly detected (TS $>$ 25), we then use the {\tt gtfindsrc} tool to refine its localization and obtain a best-fit position. This best-fit test source is subsequently added to the background model, and the likelihood fitting process is rerun with the updated model. During the temporal analysis, low-intensity background sources (TS $<$ 10) are excluded from the model. Flux and spectral extraction are performed using an {\tt unbinned} likelihood analysis with the {\tt gtlike} task\footnote{\url{https://fermi.gsfc.nasa.gov/ssc/data/analysis/scitools/likelihood_tutorial.html}}. In cases where the detection is tentative, the {\tt pyLikelihood UpperLimits} tool is employed to calculate a 95\%  confidence-level(C.L.) upper limit instead of a flux.

\subsection{Fermi-LAT Result} \label{subsec:2}

First, we analyzed the entire 16.4 years of \emph{Fermi}-LAT data to search for possible $\gamma$-ray emission from CSOs that had not been previously detected. For each source, we assumed a point source located at the radio position and adopted a power-law spectrum with the photon index fixed at $\Gamma = 2.0$. No significant excess emission above 100~MeV (TS $> 25$) was found for JVAS~J1311+1658 or any of the other previously undetected CSOs. 

\begin{figure*}
    \centering
    \includegraphics[scale=0.4]{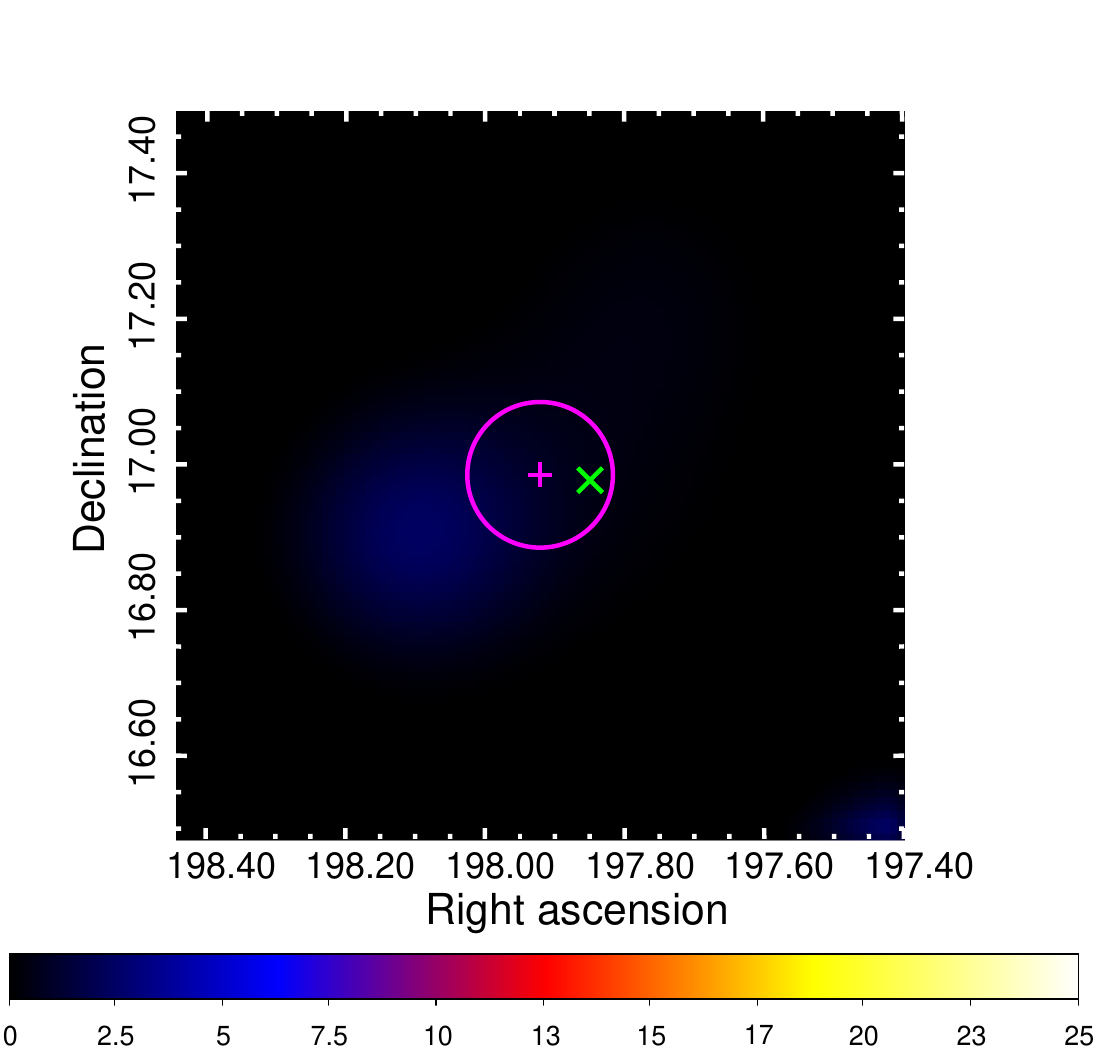}
    \includegraphics[scale=0.4]{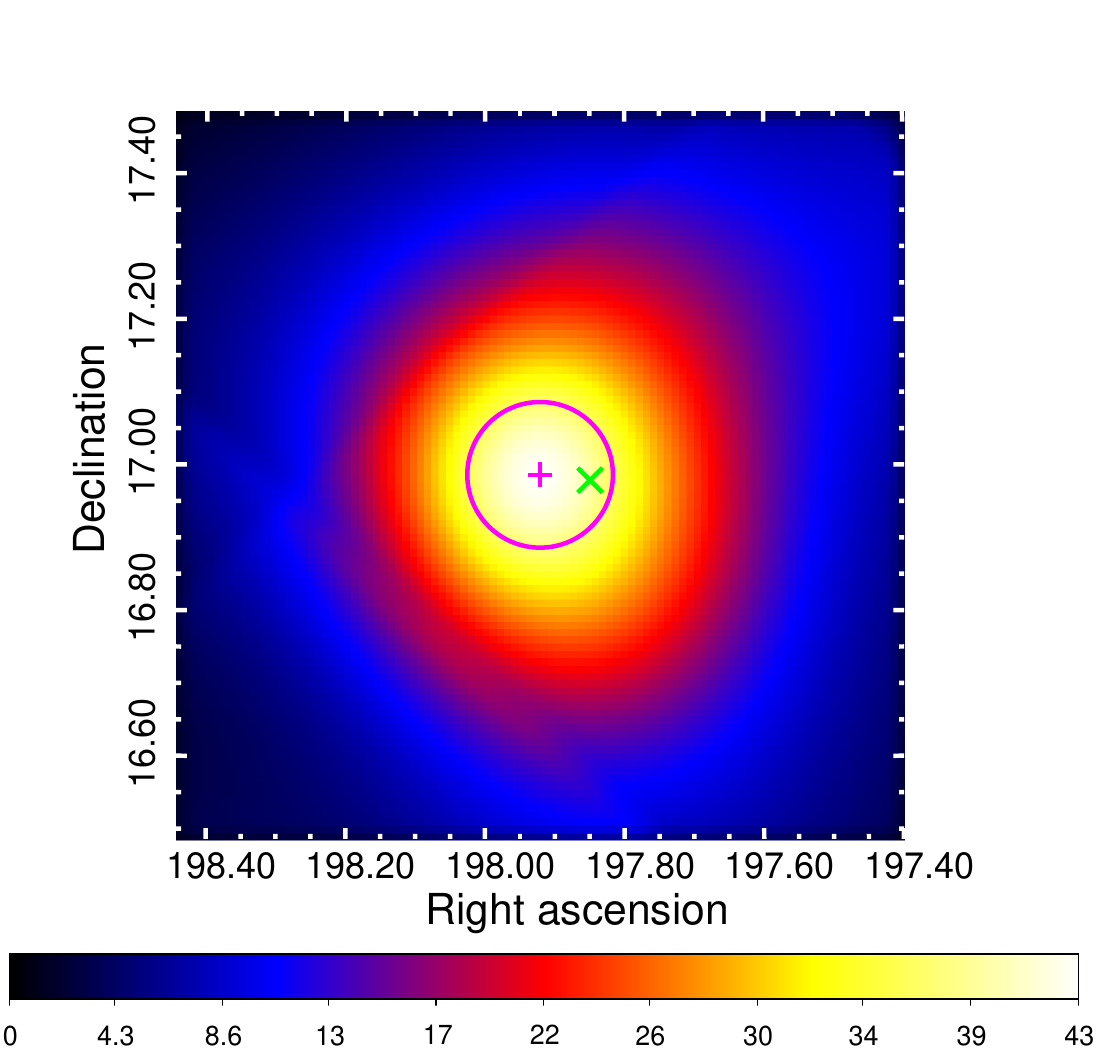}
    \includegraphics[scale=0.5]{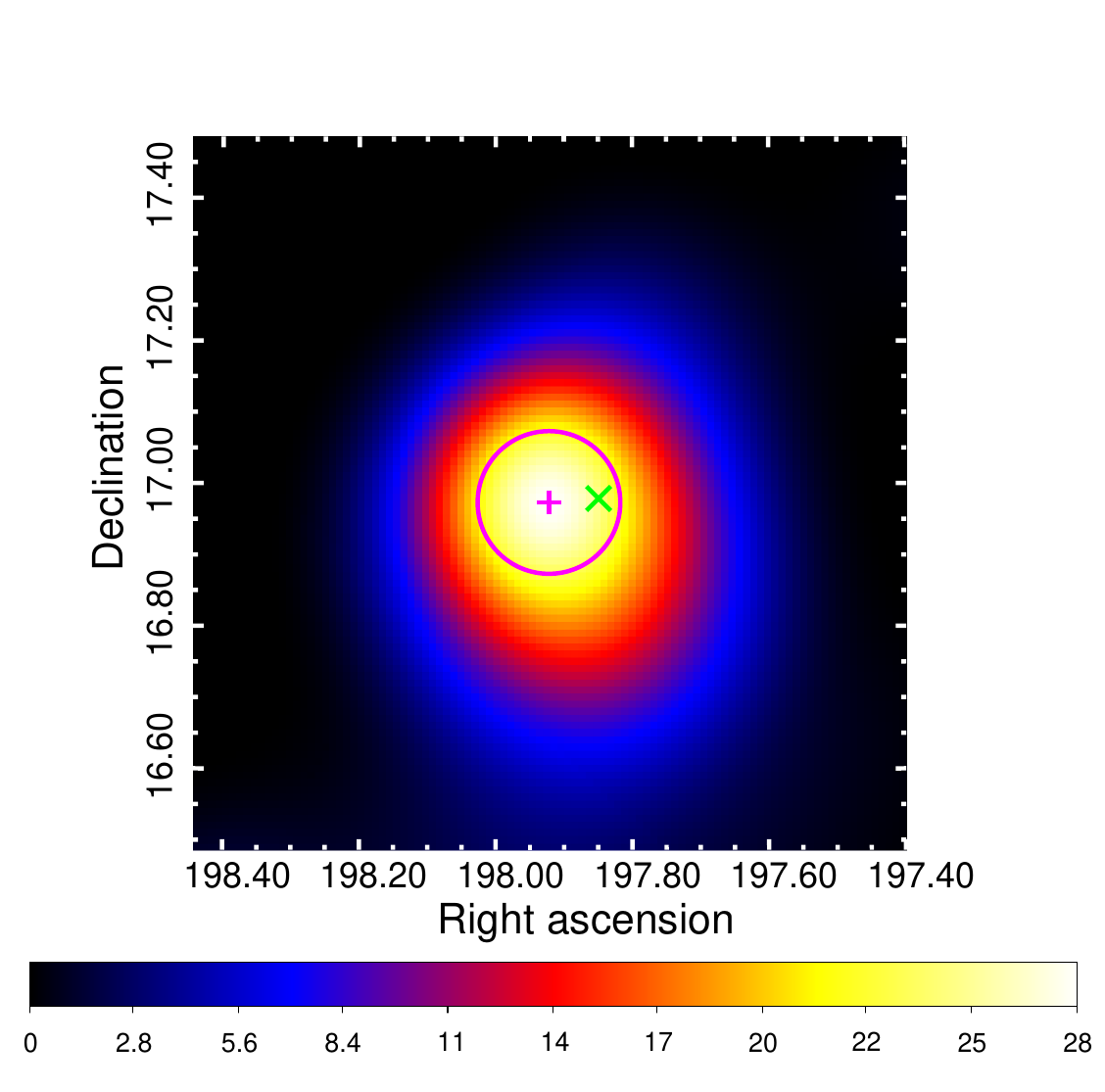}
    \caption{
    {\bf Upper panels:} Smoothed 100~MeV--300~GeV $\gamma$-ray TS maps of a
    $1^{\circ} \times 1^{\circ}$ region with a pixel size of $0.01^{\circ}$,
    where the target source is excluded from the background model.
    {\it Left panel:} TS map derived from the first 14 years of \emph{Fermi}-LAT data
    (MJD 54683--59796), during which no significant $\gamma$-ray emission is detected.
    {\it Right panel:} TS map corresponding to the 100-day $\gamma$-ray flaring interval
    (MJD 60032.6--60132.6).
    The magenta cross and circle indicate the best-fit position of the $\gamma$-ray source
    and its 95\% confidence-level localization uncertainty, respectively.
    The green cross marks the radio position of JVAS~J1311+1658.
    {\bf Bottom panel:} Same as the upper right panel, but showing the TS map in the
    1--300~GeV energy range for the flaring interval (MJD 60032.6--60132.6).}

    \label{tsmap}
\end{figure*}

\begin{figure*}
    \centering
    \includegraphics[scale=0.8]{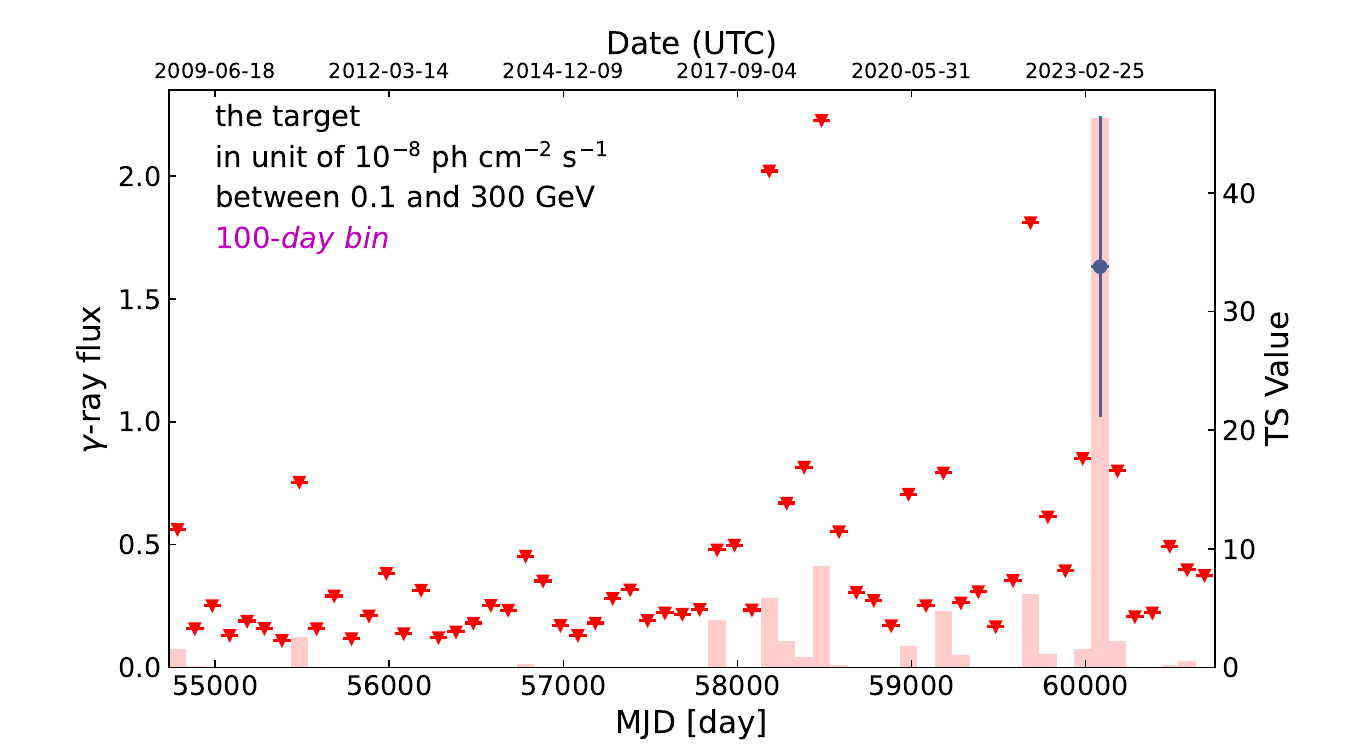}

    \begin{minipage}{0.48\textwidth}
        \centering
        \includegraphics[width=\textwidth]{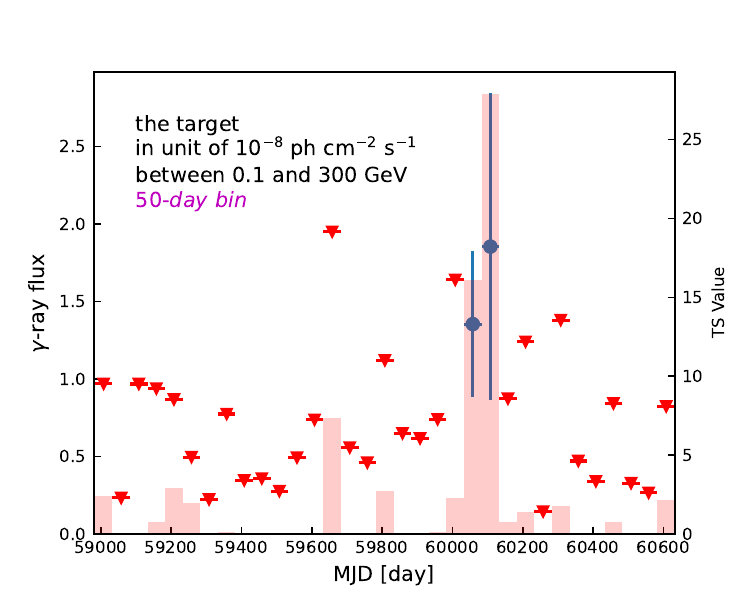}
    \end{minipage}
    \hfill
    \begin{minipage}{0.48\textwidth}
        \centering
        \includegraphics[width=\textwidth]{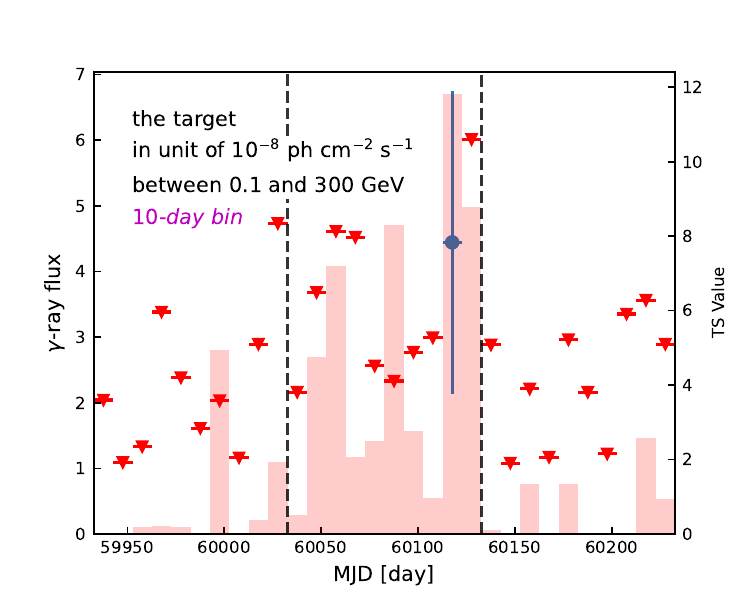}
    \end{minipage}

    \caption{
    {\bf Upper panel:} $\gamma$-ray light curve of the target source obtained with
    100-day time bins, covering the period from MJD 54733 to MJD 60683
    (approximately 16.4 years of \emph{Fermi}-LAT observations).
    Blue circles denote the measured $\gamma$-ray fluxes, red triangles indicate
    95\% C.L. flux upper limits (TS $<10$), and the semi-transparent red bars
    represent the corresponding TS values.
    {\bf Bottom panels:}
    {\it Left:} zoomed-in view of the same light curve using 50-day time bins.
    {\it Right:} zoomed-in view using 10-day time bins.
    The time interval marked by the black dashed lines corresponds to the
    100-day $\gamma$-ray flaring episode.
    }
    \label{lightcurve}
\end{figure*}

Next, we generated a $\gamma$-ray light curve with a time bin of 100 days to search for possible activity that might be hidden by background emission.
We detect a significant $\gamma$-ray flare at the position of JVAS~J1311+1658 during MJD~60032.6--60132.6, with a test statistic of $\mathrm{TS}=43$, corresponding to a detection significance of $\sim6.2\sigma$.
In contrast, no significant emission is detected during the first 14 years of \emph{Fermi}-LAT observations accumulated between MJD~54683 and 59796, with a corresponding $\mathrm{TS} = \mathrm{2}$ and a derived 95\% C.L. upper limit on the photon flux of $5 \times 10^{-10}\ \mathrm{ph\ cm^{-2}\ s^{-1}}$, which naturally explains its absence from the 4FGL-DR4 catalog.
The corresponding TS maps are shown in Figure~\ref{tsmap}, where the left panel corresponds to the non-detection period and the right panel to the flaring interval, and the 100-day binned light curve is shown in the first panel of Figure~\ref{lightcurve}.
The best-fit $\gamma$-ray source position is consistent with JVAS~J1311+1658, and a cross-check with the NRAO VLA Sky Survey (NVSS; \citealt{1998AJ....115.1693C}) and the Radio Fundamental Catalog reveals no other radio sources within the 95\% confidence localization region. This strongly suggests that the detected $\gamma$-ray emission is associated with JVAS~J1311+1658.

To further investigate the $\gamma$-ray variability timescale, we divided the 100-day flaring interval into two 50-day time bins. The flux levels in the two bins are comparable, both at the level of $\sim 10^{-8}\ \mathrm{ph\ cm^{-2}\ s^{-1}}$, with corresponding TS values of 16.1 and 27.8, respectively. We also constructed a light curve with 10-day time bins during the flaring epoch, but found no evidence for significant variability on timescales of several days. The 50-day and 10-day binned light curves are shown in the second and third panels of Figure~\ref{lightcurve}, respectively.

\begin{figure*}
    \centering
    \includegraphics[scale=0.8]{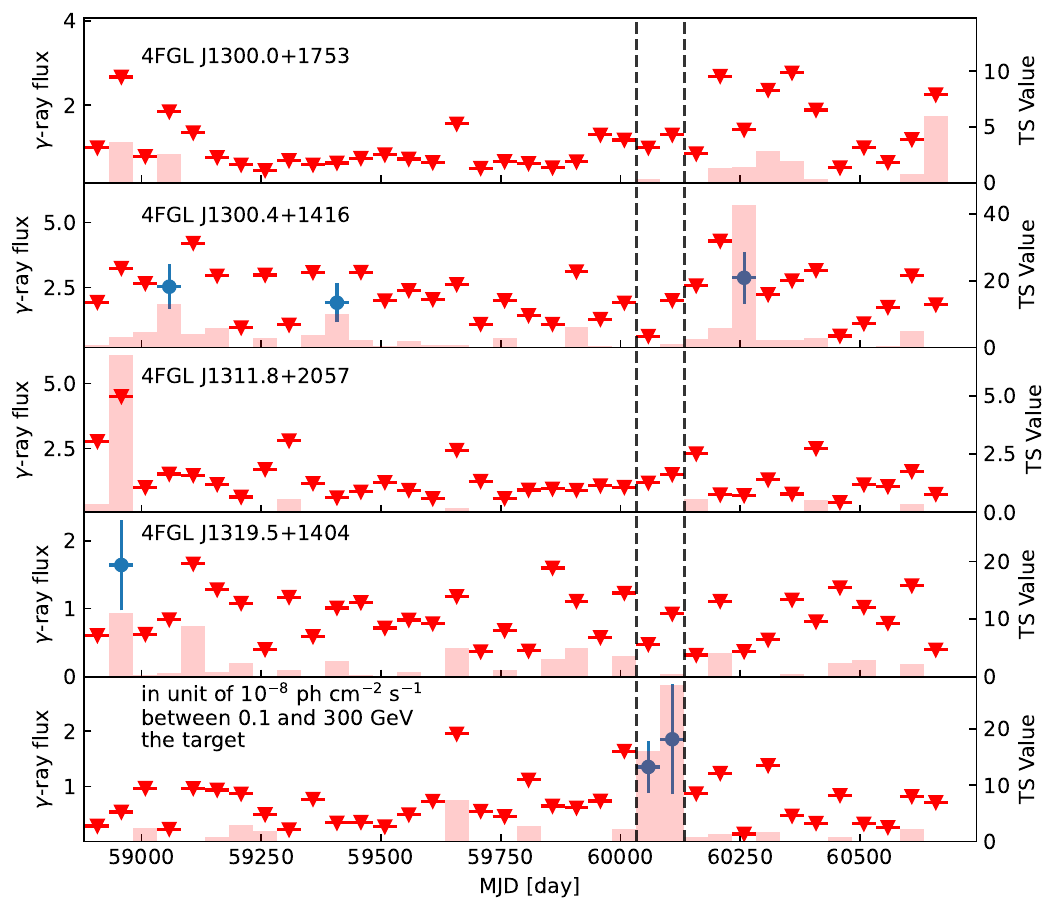}
    \caption{A zoomed-in view of the 50-day binned $\gamma$-ray light curve, shown together with several known Fermi-LAT sources in the vicinity of the target.
}
    \label{lightcurve2}
\end{figure*}

At low energies, particularly around 100~MeV, the \emph{Fermi}-LAT point-spread function (PSF) is relatively broad, with a 68\% containment radius of approximately $5^\circ$ \citep{2020ApJS..247...33A}. To assess whether the detected emission could be contaminated by nearby $\gamma$-ray sources, we compared the light curve of JVAS~J1311+1658 with those of neighboring 4FGL-DR4 sources located within the LAT PSF. As shown in Figure~\ref{lightcurve2}, JVAS~J1311+1658 exhibits a pronounced flaring state during MJD 60032.6--60132.6, whereas the nearby sources remain in relatively low $\gamma$-ray states and show no correlated variability, strongly disfavoring contamination from surrounding sources.
To further suppress possible contributions from low-energy photons with large positional uncertainties, we repeated the likelihood analysis in the 1--300~GeV energy range. During the same flaring period, a significant detection is still obtained with TS $= 28$ and a photon flux of $(8.9 \pm 3.5) \times 10^{-10}\ \mathrm{ph\ cm^{-2}\ s^{-1}}$. The best-fit position is R.A.~$=197.922^\circ$ and Dec.~$=16.9734^\circ$, with a 95\% C.L. localization uncertainty of $0.095^\circ$, consistent with that derived from the 0.1--300~GeV analysis. JVAS~J1311+1658 remains well within this error region. The corresponding TS map is shown in the third panel of Figure~\ref{tsmap}.

\begin{figure*}
    \centering
    \includegraphics[scale=0.8]{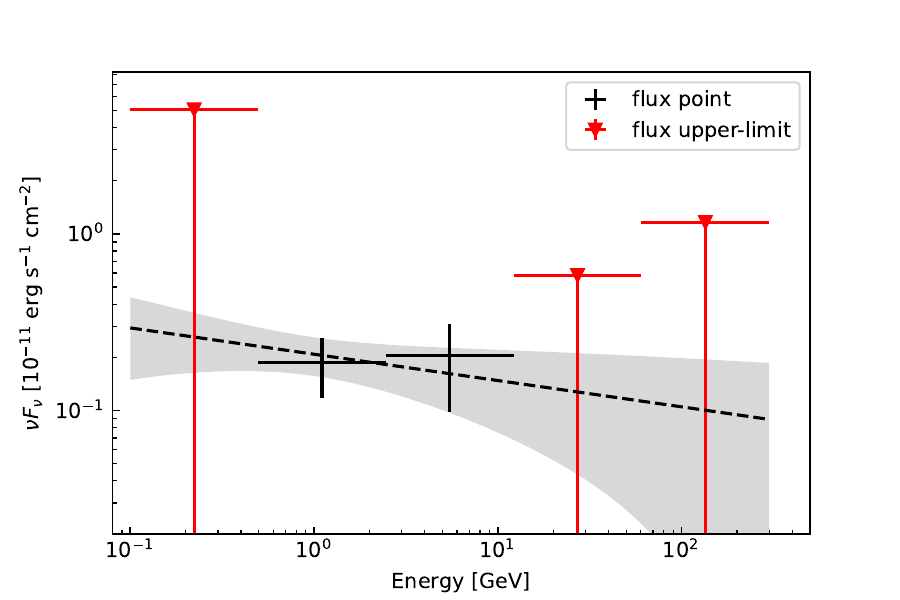}
    \caption{The $\gamma$-ray spectral energy distribution (SED) of the target source in the 100 MeV–300 GeV range. The black points represent the energy flux, the red triangles indicate the upper limits (TS $<$ 10), the black dashed line shows the best-fit model, and the gray shaded area denotes the 1$\sigma$ uncertainty range.
}
    \label{gamma_SED}
\end{figure*}

The $\gamma$-ray emission from JVAS~J1311+1658 was modeled using a power-law spectrum of the form
\begin{equation}
\frac{dN}{dE} = N_0 \left( \frac{E}{E_0} \right)^{-\Gamma}.
\end{equation}
For the entire 100-day flaring period, spectral fitting over the 0.1--300~GeV energy range yields a photon index of $\Gamma = 2.15 \pm 0.19$ and a normalization of
$N_0 = (1.30 \pm 0.32) \times 10^{-12}\ \mathrm{MeV^{-1}\ cm^{-2}\ s^{-1}}$
at a reference energy of $E_0 = 1000$~MeV. The corresponding average photon flux is
$(1.60 \pm 0.61) \times 10^{-8}\ \mathrm{ph\ cm^{-2}\ s^{-1}}$,
which translates to an energy flux of
$(1.41 \pm 0.34) \times 10^{-11}\ \mathrm{erg\ cm^{-2}\ s^{-1}}$.
Dividing the flaring interval into two 50-day bins yields consistent spectral properties, with comparable $\gamma$-ray fluxes at the level of $\sim 10^{-8}\ \mathrm{ph\ cm^{-2}\ s^{-1}}$. The corresponding photon indices are $\Gamma = 2.12 \pm 0.24$ and $\Gamma = 2.20 \pm 0.27$, respectively.
In addition, we performed a spectral analysis over the 1--300~GeV band for the full 100-day interval, obtaining a normalization of
$N_0 = (0.89 \pm 0.51) \times 10^{-12}\ \mathrm{MeV^{-1}\ cm^{-2}\ s^{-1}}$.
The photon index is $\Gamma = 1.99 \pm 0.37$, which is slightly harder than that derived from the broader 0.1--300~GeV energy range, but consistent within uncertainties. This fit yields an average photon flux of
$(8.90 \pm 3.47) \times 10^{-10}\ \mathrm{ph\ cm^{-2}\ s^{-1}}$,
corresponding to an energy flux of
$(8.96 \pm 7.04) \times 10^{-12}\ \mathrm{erg\ cm^{-2}\ s^{-1}}$.
A summary of the spectral parameters and TS values for JVAS~J1311+1658, together with other $\gamma$-ray--emitting CSOs, is provided in Table~\ref{table1}. The $\gamma$-ray spectral energy distribution shown in Figure~\ref{gamma_SED} indicates that the emission is primarily concentrated around a few GeV, with negligible contribution at energies above several tens of GeV.

Finally, we employed the {\tt gtsrcprob} tool to estimate the probability that individual $\gamma$-ray photons are associated with JVAS~J1311+1658. The detected emission is dominated by four photons with association probabilities exceeding 90\%, including one photon at 1.3~GeV and three photons at 4.3, 6.8, and 6.9~GeV. This further supports the conclusion that JVAS~J1311+1658 lacks significant very-high-energy $\gamma$-ray emission.

\subsection{Other Observations} \label{subsec:3}

The redshift of JVAS J1311+1658 was first reported by \citet{2016MNRAS.459..820T} as a photometric redshift of $z = 0.03$ based on data from the Sloan Digital Sky Survey (SDSS; \citep{2000AJ....120.1579Y}). Later, \citet{2024ApJ...961..240K}, referencing the 13th Data Release of the SDSS \citep{2017ApJS..233...25A}, provided an updated spectroscopic redshift of $z = 0.0814$ for the source. However, upon re-examining the 13th Data Release of the SDSS, we found no known SDSS object located at the radio position of JVAS J1311+1658, with the nearest cataloged source being about 6 arcseconds away. Furthermore, we checked the most recent SDSS Data Release 19 data \footnote{\url{https://www.sdss.org/dr19/data_access/get_data/}} as well as the newly released Dark Energy Spectroscopic Instrument (DESI; \citep{2019AJ....157..168D}) Data Release 10 data\footnote{\url{https://www.legacysurvey.org/dr10/}} and obtained the same result, indicating that the reported redshift of JVAS J1311+1658 is incorrect \footnote{We have contacted the authors of \cite{2024ApJ...961..240K}, who confirmed that the redshifts of JVAS J1311+1658 and another CSO, JVAS J0855+5751, were incorrectly reported in their paper.}.

JVAS J1311+1658 has been extensively studied in the radio band \citep{1991ApJS...75....1B, 1992ApJS...79..331W, 2002ApJS..141...13B, 2007ApJ...658..203H, 2017ApJ...836..174C}. The observed turnover frequency of the source is 0.447 GHz, and the corresponding flux density is 0.824 Jy \citep{2017ApJ...836..174C}. Based on Very Long Baseline Array (VLBA) observations, JVAS J1311+1658 was initially classified as a CSO candidate \citep{2000ApJ...534...90P, 2005ApJ...622..136G}, with the source showing a central core, a jet extending to the northwest, and a faint counterjet or hotspot to the southeast. Due to the presence of flat-spectrum emission at the central position and two edge-dimmed steep spectrum jets, the source was later formally classified as a CSO \citep{2016MNRAS.459..820T}. Recently, \citet{2024ApJ...961..240K} classified JVAS J1311+1658 as a bona fide CSO based on a new set of criteria: (i) a projected radio structure length $< 1$~kpc, (ii) detection of radio emission on both sides of the center of activity, (iii) no fractional variability greater than $20\%~\mathrm{yr}^{-1}$, and (iv) no detection of superluminal motion exceeding $v_{\mathrm{app}} = 2.5c$ in any jet component. In their latest measurements, the angular size of the source was found to be 27.0 mas in its 5 GHz radio image, corresponding to a projected linear size of $ \sim 41$~pc assuming a redshift of 0.0814.

\cite{2018ApJS..239...20M} investigated the infrared(IR) to optical/ultraviolet (UV) data of JVAS J1311+1658\footnote{\url{http://www.gaoran.ru/english/as/ac_vlbi/ocars.txt}}  and found that the source was not detected in several  catalogs that have a large number of objects, including the Wide-field Infrared Survey Explorer (WISE), the Two Micron All-Sky Survey (2MASS; \cite{2006AJ....131.1163S}), Pan-STARRS (PS1; \cite{2016arXiv161205560C}), and Gaia \citep{2016A&A...595A...2G}. Furthermore, no UV counterpart was identified in the Galaxy Evolution Explorer (GALEX) catalog \citep{2017ApJS..230...24B}, suggesting that the IR and optical/UV emission from JVAS J1311+1658 is extremely faint. At higher energies, the latest release of the RFC found that this source is associated with an X-ray source, \text{1eRASS J131123.3+165845} \citep{2025ApJS..276...38P}. The flux of this eROSITA X-ray source in the 0.2--2.3 keV band is \( (4.24 \pm 1.94) \times 10^{-14}~\mathrm{erg\,cm^{-2}\,s^{-1}} \), indicating a relatively weak flux level \citep{2024A&A...682A..34M}.

Figure \ref{multband_SED} shows the broadband SED of CSOs with gamma-ray detections. It can be seen that TXS 0128+554, NGC 3894, and NGC 6328 exhibit prominent emission from the near-IR to ultraviolet bands, which  mainly originates from the contribution of their host galaxies \citep{2022ApJ...941...52S, 2024A&A...684A..65B}. DA 362 and 4C +39.23B also show emission in the near-IR band but are extremely faint in the optical to UV bands, suggesting significant dust obscuration. However, JVAS J1311+1658 shows extremely weak emission from the IR to the UV bands, indicating that its host galaxy environment may be significantly different from that of other CSOs, with a potentially low star formation rate \citep{2003A&A...410...83H,2012MNRAS.426..330D}.

\section{Discussion and conclusion} \label{sec:4}

Our results indicate that JVAS J1311+1658, as a rare young radio galaxy, experienced a significant $\gamma$-ray outburst, underscoring the importance of identifying the origin of its emission. However, since its redshift remains uncertain, it is challenging to discuss its physical properties. Therefore, we will employ specific methods to estimate its redshift and further investigate the potential origin of its $\gamma$-ray emission.

\begin{figure*}
    \centering
    \includegraphics[scale=0.5]{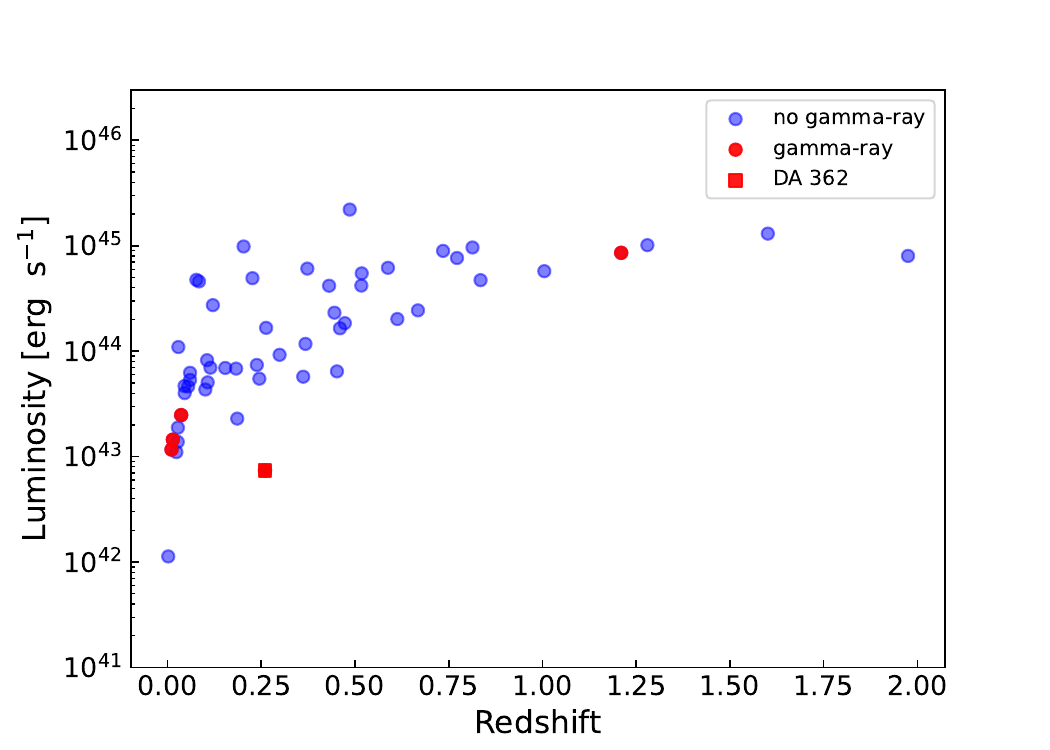}
    \includegraphics[scale=0.5]{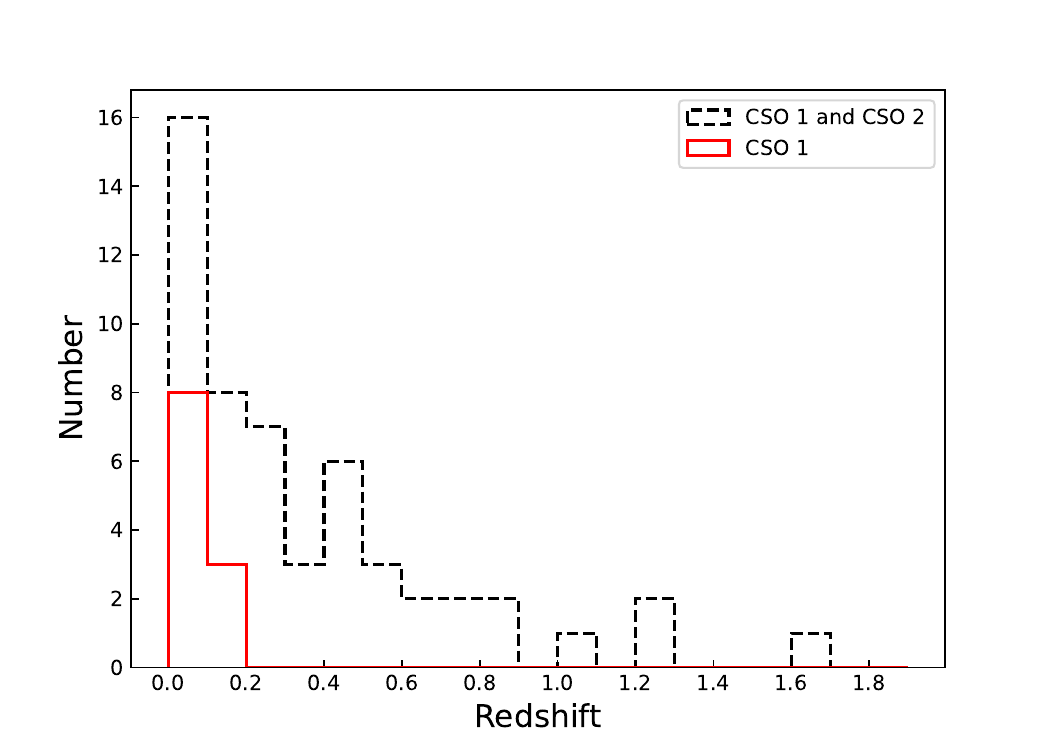}
    \caption{{\bf Left panel:} Redshift versus WISE W1-band luminosity diagram. The red points represent CSOs with detected $\gamma$-ray emission, while the blue points denote those without $\gamma$-ray detection. Except for DA 362, the redshifts of the CSOs are taken from the spectroscopic measurements reported by \citep{2024ApJ...961..240K, 2024ApJ...961..242R}. However, since the redshifts of JVAS J1311+1658 and JVAS J0855+5751 are incorrect, these two sources are excluded from the plot. For comparison, DA 362 is included, although the reliability of its redshift remains uncertain.
    {\bf Right panel:} Redshift distribution of CSOs. The red histogram represents the redshift distribution of CSO 1s, while the black histogram shows the overall distribution of all CSOs (including CSO 1 and CSO 2). All data are taken from \cite{2024ApJ...961..242R}. Similarly, JVAS J1311+1658 and JVAS J0855+5751 are excluded from the plot due to their incorrect redshift values.
}
    \label{redshift}
\end{figure*}

The left panel of Figure~\ref{redshift} shows the luminosities of all CSOs with known redshifts in the WISE W1 band. The redshifts of these CSOs range from $0$ to $2$, and their IR luminosities at 3.4 $\mu\mathrm{m}$ span from $\sim 10^{42} \, \mathrm{erg\, s^{-1}}$ to about $10^{45} \, \mathrm{erg\, s^{-1}}$, covering nearly four orders of magnitude. Here, we assume that the $3.4 \,\mu\mathrm{m}$ luminosity of JVAS J1311+1658 is as low as $10^{41} \, \mathrm{erg \, s^{-1}}$. 
In the of JVAS J1311+1658, the ALLWISE W1-band non-detection limit is approximately $17.8$~mag, corresponding to a flux density upper limit of
$S_{W1} = 306.682 \times 10^{-M_{W1}/2.5} \, \mathrm{Jy}$ \citep{2014MNRAS.439..545C}
$\approx 306.682 \times 10^{-17.8/2.5} \, \mathrm{Jy} \approx 23 \,\mu\mathrm{Jy}$.
Here, we conservatively assume that the 3.4~$\mu$m luminosity of
JVAS~J1311+1658 is as low as $\sim 10^{42}$~erg~s$^{-1}$.
Under this assumption, we derive a lower limit on the redshift of
$z_{\rm min} \approx 0.13$ for JVAS~J1311+1658.

It is well known that CSO samples are affected by significant selection effects, particularly a redshift-dependent bias. As discussed in \cite{2024ApJ...961..242R}, CSO 1 sources, which are edge-dimmed and intrinsically low-luminosity, are predominantly observed at low redshifts, whereas CSO 2 sources, which are edge-brightened and high-luminosity, can be detected over a broader redshift range. As shown in the right panel of Figure~\ref{redshift}, all previously known CSO 1 sources have redshifts below 0.2. This is largely due to their low surface brightness, which makes high-redshift CSO 1s difficult to detect in existing flux-limited surveys. JVAS~J1311+1658 is classified as a CSO 1 source. We adopt a conservative upper limit of $z \sim 0.2$ for its redshift. Considering its WISE W1-band non-detection, which provides a lower limit of $z \gtrsim 0.13$, we thus adopt a plausible redshift range of $0.13 \lesssim z \lesssim 0.2$ for this source. 
This assumption enables us to estimate its luminosities and compare them with those of other CSOs
and young radio sources.

\begin{figure*}
    \centering
    \includegraphics[scale=0.8]{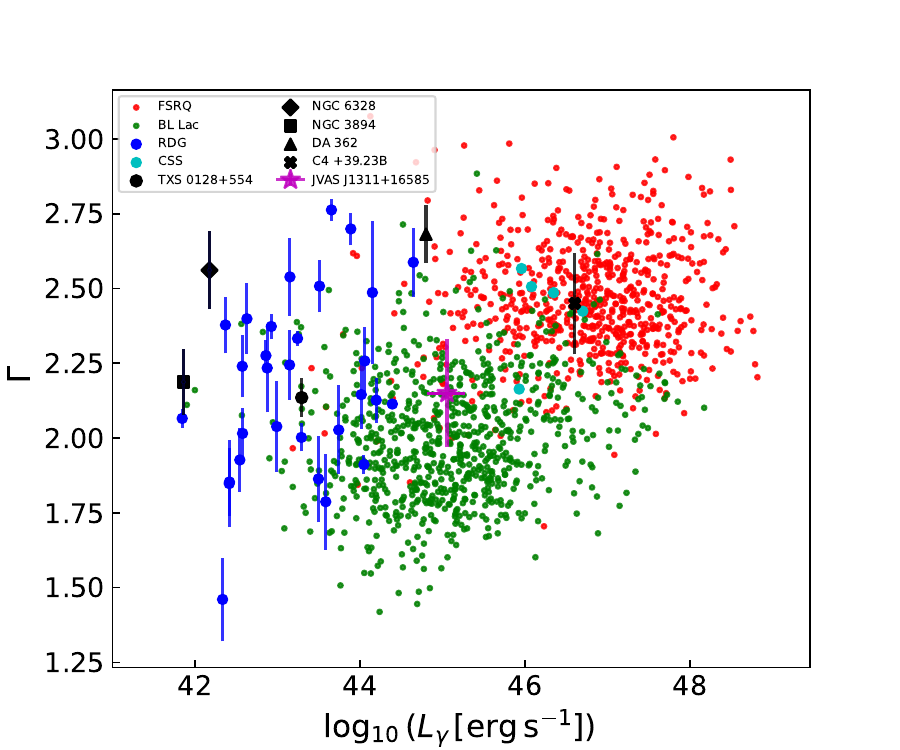}
    \caption{A plot of $\gamma$-ray luminosity versus $\gamma$-ray photon index for jetted AGNs with known redshifts included in the 4LAC-DR3. The jetted AGNs are represented by different symbols: FSRQs (red points), BL Lacs (green points), radio galaxies (blue points), and compact steep-spectrum sources (CSS; cyan points). The data for 4C +39.23B are adopted from \cite{2022ApJ...927..221G}, while the results for JVAS J1311+16585 are derived from our 100-day analysis of Fermi-LAT data spanning MJD 60032.6 to 60132.6. Considering that the redshift of JVAS J1311+16585 is constrained to the range 0.13 – 0.2, its luminosity range is indicated by a magenta star with error bars in the figure. All $\gamma$-ray luminosities in the plot have been corrected for K-correction.}.

    \label{lum_vs_index}
\end{figure*}

To investigate the origin of $\gamma$-ray emission in JVAS J1311+1658, we compared its properties with those of other extragalactic $\gamma$-ray sources. Figure \ref{lum_vs_index}  presents the distribution of $\gamma$-ray luminosity versus photon spectral index (i.e., the $L_{\gamma}$--$\Gamma$ plane) for a selected sample of $\gamma$-ray–detected AGNs,
including flat-spectrum radio quasars (FSRQs), BL Lac objects (BL Lacs), CSOs, radio galaxies, and compact steep-spectrum sources (CSSs),
drawn from the 4LAC-DR3 catalog
\footnote{\url{https://fermi.gsfc.nasa.gov/ssc/data/access/lat/4LACDR3/}}. As expected, FSRQs and BL Lac objects exhibit the highest $\gamma$-ray luminosities, reflecting strong Doppler-boosting effects from highly relativistic jets aligned close to the line of sight \citep{2017A&ARv..25....2P,2019ARA&A..57..467B}. FSRQs typically display softer spectra and higher luminosities than BL Lacs, implying different $\gamma$-ray emission mechanisms: in BL Lacs, $\gamma$-rays are mainly produced via synchrotron-self-Compton (SSC) emission \citep{1996ApJ...461..657B}, whereas in FSRQs they arise from external-Compton (EC) scattering of photons from external sources \citep{1994ApJ...421..153S}.
In addition, we consider CSSs, which represent a more evolved class of young radio sources compared to CSOs. Their radio structures are typically scaled-down versions of those of Fanaroff–Riley type II galaxies and are considered
progenitors of extended radio galaxies \citep{1996ApJ...460..612R,
2014MNRAS.438..463O}. As shown in Figure~\ref{lum_vs_index}, the CSS sources in our selected sample occupy a region of $\gamma$-ray luminosity comparable to that of
FSRQs, indicating that relativistic jets and Doppler beaming may play an important role in their high-energy emission, consistent with previous studies
\citep{2021MNRAS.507.4564P}. Since CSSs are expected to evolve into extended radio galaxies, it is plausible that the strength of relativistic beaming diminishes as the jets expand, potentially leading to lower $\gamma$-ray luminosities in more
extended sources.

In contrast, classical radio galaxies generally occupy the low-luminosity regime in the $L_\gamma$–$\Gamma$ plane, indicating minimal contribution from Doppler-enhanced jets. The $\gamma$-ray luminosities of both NGC 6328 and NGC 3894 are very low, consistent with those of typical radio galaxies. TXS 0128+554 has a higher $\gamma$-ray luminosity than the former two, placing it near the region of low-luminosity BL Lac objects while still within the radio galaxy domain. The spectral index of JVAS J1311+1658 is comparable to that of TXS 0128+554 and NGC 3894, but its $\gamma$-ray emission is much stronger, exhibiting characteristics of either BL Lac objects or the most luminous radio galaxies. This strong $\gamma$-ray emission from JVAS J1311+1658 is likely associated with an enhanced activity state near the jet base, related to its recently launched jets, as proposed by \cite{2020ApJ...899..141L} in their discussion of TXS 0128+554. Although DA 362 is located near the highest-luminosity radio galaxies, it should be noted that the reliability of its redshift is unclear, and thus its $\gamma-$ray luminosity is highly uncertain \citep{2025ApJ...979...97S}. 
In contrast, 4C+39.23B exhibits an exceptionally high $\gamma$-ray luminosity, placing it well outside the region typically occupied by radio galaxies in Figure~\ref{lum_vs_index} and closer to the parameter space populated by FSRQs. This behavior suggests that Doppler-boosted jet emission may play an important role in its $\gamma$-ray output. One possible explanation is that the high $\gamma$-ray luminosity originates from a compact region within the inner jet that is locally or temporarily oriented closer to the line of sight, analogous to the scenario proposed for 3C~120 \citep{2015ApJ...808..162C}.
However, $\gamma$-ray emission dominated by strong Doppler boosting is generally expected to be accompanied by detectable variability on timescales of months. In the case of 4C+39.23B, no significant $\gamma$-ray variability is observed over its $\sim$3-year detection period \citep{2022ApJ...927..221G}, which places constraints on a strongly Doppler-boosted and transient origin of the emission.
In addition, \citet{2022ApJ...927..221G} show that the angular separation between the $\gamma$-ray emission attributed to 4C+39.23B and the nearby bright blazar 4FGL~J0824.9+3915 (associated with 4C+39.23A) is only $\sim$0.1$^\circ$.
 Their analysis further indicates that the $\gamma$-ray signal associated with 4C+39.23B is predominantly detected at energies below 1~GeV, where the \textit{Fermi}-LAT PSF is relatively broad. Based on these observational facts, we note that PSF leakage from the nearby blazar cannot be excluded.

For radio galaxies with large-scale jets, $\gamma$-ray emission can originate from their extended radio lobes. This has been conclusively demonstrated by \cite{2010Sci...328..725A} and \cite{2016ApJ...826....1A}, who unambiguously detected GeV-band emission from the extended lobes of Centaurus A (Cen A) and Fornax A. Similarly, for CSOs, \cite{2008ApJ...680..911S} and \cite{2010ApJ...715.1071O} proposed that $\gamma$-ray emission could result from inverse Compton (IC) scattering of ambient low-energy photons by nonthermal electrons within the radio lobes. More recently, \citet{2022ApJ...941...52S} successfully explained the $\gamma$-ray emission from NGC 6328 as IC scattering of UV photons—presumably from an accretion flow—by nonthermal electrons in its expanding radio lobes.

\begin{figure*}
    \centering
    \includegraphics[scale=0.7]{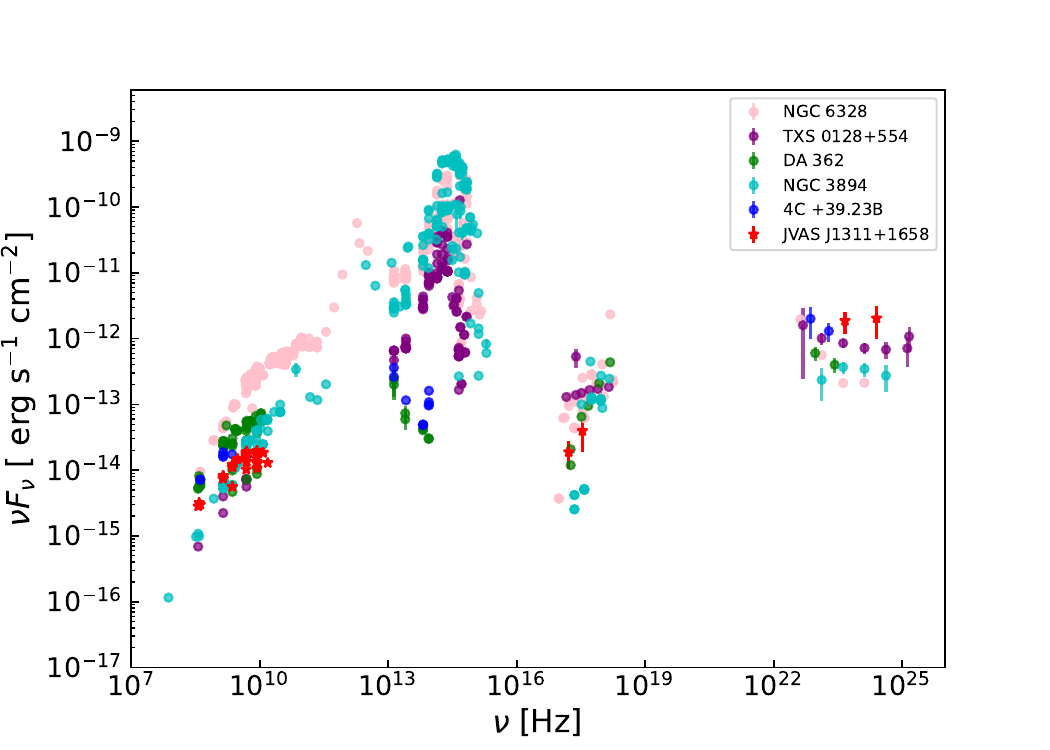}
    \caption{Multi-band SEDs of the six CSOs with detected $\gamma$-ray emission. Most of the data are collected from archival observations provided by the Space Science Data Center (SSDC; \url{https://tools.ssdc.asi.it/SED/}). For DA 362, the X-ray data are taken from \cite{2025ApJ...979...97S}, and the $\gamma$-ray data are obtained from our 14-year analysis of Fermi-LAT observations. The $\gamma$-ray data for 4C +39.23B are adopted from \cite{2022ApJ...927..221G}.
}
    \label{multband_SED}
\end{figure*}

Assuming that the $\gamma$-ray emission of JVAS~J1311+1658 is produced via IC scattering of accretion-disk UV photons by relativistic electrons in its young radio lobes, we estimate the expected $\gamma$-ray luminosity following the approach of \citep{2008ApJ...680..911S,2021MNRAS.507.4564P}:
\begin{equation}
\label{eq2}
\frac{(\epsilon L_{\epsilon})_{\rm IC/UV}}{10^{42}~{\rm erg~s^{-1}}} \sim 2  \frac{\eta_e}{\eta_B}
\left( \frac{L_j}{10^{45}~{\rm erg~s^{-1}}} \right)^{1/2}
\left( \frac{LS}{100~{\rm pc}} \right)^{-1}
\left( \frac{L_{\rm UV}}{10^{46}~{\rm erg~s^{-1}}} \right)
\left( \frac{\epsilon}{1~{\rm GeV}} \right)^{-0.25}~,
\end{equation}
where $L_j$ is the jet power, $L_{\rm UV}$ is the UV luminosity, and $\eta_e/\eta_B$ denotes the ratio between the electron and magnetic-field energy densities.
Assuming a redshift of $z=0.2$, the projected linear size is $LS \sim 100$~pc. Following model~A1 of \citet{2022ApJ...941...52S}, we conservatively adopt $\eta_e/\eta_B \sim 10$ and $L_{\rm UV} \sim 10^{43}$~erg~s$^{-1}$. The jet power can be estimated as \citep{2020ApJ...892..116W}:
\begin{equation}
L_j \sim 1.5 \times 10^{45}
\left( \frac{LS}{100~{\rm pc}} \right)^{9/7}
\left( \frac{\tau_j}{100~{\rm yr}} \right)^{-1}
\left( \frac{L_{\rm 5~GHz}}{10^{42}~{\rm erg~s^{-1}}} \right)^{4/7}
~{\rm erg~s^{-1}} ,
\end{equation}
To avoid violating the CSO definition requiring the apparent speed to remain below 
$v_{\rm app}=2.5\,c$, we assume a source age of $\tau_j \sim 200$~yr. 
Adopting a 5~GHz flux of $F_{\rm 5~GHz} \sim 2 \times 10^{-14}$~erg~cm$^{-2}$~s$^{-1}$ (see Figure~\ref{multband_SED}), the corresponding radio luminosity is $L_{\rm 5~GHz} \sim 2.5 \times 10^{42}$~erg~s$^{-1}$, which yields a jet power of $L_j \sim 10^{45}$~erg~s$^{-1}$.
Combining with Equation~\ref{eq2}, the predicted GeV $\gamma$-ray luminosity is 
$(\epsilon L_{\epsilon})_{\rm IC/UV} \sim 2 \times 10^{40}$~erg~s$^{-1}$, 
corresponding to an expected energy flux of 
$\sim 2 \times 10^{-16}$~erg~cm$^{-2}$~s$^{-1}$, 
which is approximately four orders of magnitude lower than the 
$\gamma$-ray flux inferred from the \emph{Fermi}-LAT observations
\footnote{The adopted ratio $\eta_e/\eta_B \sim 10$ is a conservative choice; 
a stronger deviation from energy equipartition 
(i.e., $\eta_e/\eta_B \gg 1$) would yield a higher predicted IC/UV flux.}.
In addition to accretion-disk UV photons, we also considered other potential seed photon fields for IC scattering, including the synchrotron photons produced by the relativistic electrons in the radio lobes themselves, stellar emission from the host galaxy, IR emission from a dusty torus, and the cosmic microwave background (CMB). Even under optimistic assumptions, the $\gamma$-ray luminosity produced by IC scattering off these additional photon fields is comparable to or lower than the IC/UV contribution estimated above. Therefore, regardless of the choice of plausible seed photon fields, IC emission from the young radio lobes cannot account for the observed $\gamma$-ray emission of JVAS~J1311+1658.

A similar discrepancy is found in TXS~0128+554. When an IC model for the radio-lobe electrons is applied—accounting for multiple seed photon fields including near-IR stellar emission from the host galaxy, synchrotron radiation from the lobe electrons, IR emission from the dusty torus, UV radiation from the accretion disk, and the cosmic microwave background—the predicted $\gamma$-ray flux is approximately three orders of magnitude lower than that observed by \emph{Fermi}-LAT. This strongly suggests that the $\gamma$-ray emission of TXS~0128+554 is unlikely to originate from its radio lobes \citep{2020ApJ...899..141L}. As shown in Figure~\ref{multband_SED}, the $\gamma$-ray luminosity of NGC~6328 is comparable to its radio luminosity, whereas those of TXS~0128+554 and JVAS~J1311+1658 are significantly higher than their respective radio luminosities\footnote{The $\gamma$-ray and radio data for JVAS~J1311+1658 are not simultaneous, and no radio observations cover the $\gamma$-ray flaring period, which may introduce additional uncertainty.}.

Moreover, considering the significant $\gamma$-ray variability exhibited by JVAS J1311+1658, we refer to the behavior observed in the compact structures of the radio galaxies 3C 111 and 3C 120, where superluminal knots ejected from the radio core are often accompanied by intense $\gamma$-ray flares \citep{2012ApJ...751L...3G, 2015ApJ...808..162C, 2015ApJ...799L..18T}. This suggests that the $\gamma$-ray emission in radio galaxies can also originate from the sub-pc jet or core region. Based on this reasoning, we propose that the $\gamma$-ray outbursts of JVAS J1311+1658 are also likely produced by renewed activity in its jet core as discussed above. Similarly, DA 362 exhibits $\gamma$-ray variability on monthly timescales, as observed for JVAS J1311+1658, underscoring the substantial diversity of $\gamma$-ray emission among CSOs, which suggests that the $\gamma$-ray environments of CSOs are complex and not yet fully understood. Long-term multiwavelength monitoring of JVAS J1311+1658 is crucial, as it may provide important information about the origin of its $\gamma$-ray emission. In particular, if flux enhancements are detected in other bands, such as radio or X-rays, during $\gamma$-ray flares, this would help constrain the emission site and improve our understanding of the high-energy radiation properties of such young radio sources.

In summary, we analyzed the \textit{Fermi}-LAT data and found that, during the period from MJD 60032.6 to 60132.6, a $>5\sigma$ significant $\gamma$-ray source was detected near the position of JVAS J1311+1658. This source does not appear in the latest 4FGL-DR4 catalog, nor in any of the previous \textit{Fermi}-LAT catalogs. Based on our revised best-fit $\gamma-$ray position, we obtained a TS value of 43 for this source.
The radio position of JVAS J1311+1658 is very close to the best-fit position of the $\gamma$-ray source, with a separation of approximately 0.065 degrees, and it falls within the 95$\%$ confidence-level error circle of the $\gamma$-ray source. Based on the Bayesian method of spatial coincidence, we concluded that this source is most likely the $\gamma$-ray counterpart to JVAS J1311+1658. This is the first detection of GeV $\gamma-$ray emission from the young radio galaxy JVAS J1311+1658, making it the sixth CSO confirmed as a $\gamma-$ray emitter, a relatively rare class of Fermi LAT sources. Its strong $\gamma$-ray emission and variability suggest that the $\gamma$-ray radiation most likely originates from newly launched jets at the core and it may provide important insights into the high-energy radiation observed in young radio sources and the evolution of their jets.

\section*{Acknowledgments}
We sincerely thank the anonymous referee for their valuable comments, which have led to notable improvements in this manuscript.
We acknowledge the use of publicly available Python libraries, including Astropy \citep{2013A&A...558A..33A}, Matplotlib \citep{hunter2007matplotlib}, Pandas \citep{reback2020pandas}, NumPy, and SciPy \citep{2020NatMe..17..261V}, which greatly facilitated our analysis. This work has made use of data from the High Energy Astrophysics Science Archive Research Center (HEASARC), provided by NASA’s Goddard Space Flight Center. We also utilized the NASA/IPAC Extragalactic Database (NED), operated by the Jet Propulsion Laboratory, California Institute of Technology, under contract with NASA.

This work was  supported by the National Natural Science Foundation of China (NSFC) under grants No. 12473049.

\bibliography{refs}
\bibliographystyle{aasjournal}


\begin{table}[ht]
    \centering
    \caption{Fermi-LAT Data Analysis Results for JVAS J1311+1658 and Other $\gamma$-ray-emitting CSOs.}
    \begin{tabular}{lccccc}
        \toprule
        CSO Name & 4FGL Name & \makecell[c]{$F_{0.1-300\,\mathrm{GeV}}$ \\ {\scriptsize $[10^{-8}\,\mathrm{ph\,cm^{-2}\,s^{-1}}]$}} & \makecell[c]{$\nu L_{\nu \, 0.1-300\,\mathrm{GeV}}$ \\ {\scriptsize $[10^{-12 }\,\mathrm{erg\,cm^{-2}\,s^{-1}}]$}} & $\Gamma_{\gamma}$  & TS\\ 
        \midrule
        JVAS J1311+1658   & ${}$   & $1.60 \pm 0.60$  & $14.1 \pm 3.43$ & $2.15 \pm 0.19$ & 43 \\ 
        TXS 0128+554    & 4FGL J0131.2+5547   & $0.69 \pm 0.07 $ &$5.73 \pm 0.62$  & $2.14 \pm 0.08$ & 236 \\ 
        NGC 6328 & 4FGL J1724.2-6501  & $0.55 \pm 0.13$ & $2.61 \pm 0.64$   & $2.59 \pm 0.15$ & 43   \\ 
        NGC 3894    & 4FGL J1149.0+5924 & $0.34 \pm 0.08$ & $2.52 \pm 0.57$  & $2.17 \pm 0.14$ & 118 \\
        DA 362    & 4FGL J1416.0+3443 & $0.58 \pm 0.11$ & $2.47 \pm 0.48$ & $2.67 \pm 0.11$   & 82 \\
        4C +39.23B & ${}$ & $0.94 \pm 0.41$ & $3.72 \pm 0.15$ & $2.45 \pm 0.17$   & 31 \\
        \bottomrule
    \end{tabular}
    {\footnotesize {\bf NOTE:} Except for JVAS J1311+16585 and 4C +39.23B, the results for the remaining CSOs are derived from our analysis of 14 years of Fermi-LAT data. This may explain why the flux and TS of DA 362 are lower than those reported by \cite{2025ApJ...979...97S}, who used approximately 15.75 years of Fermi-LAT observations. The results for JVAS J1311+16585 are based on our 100-day analysis of Fermi-LAT data from MJD 60032.6 to 60132.6, while those for 4C +39.23B are adopted from \cite{2022ApJ...927..221G}, who analyzed 1000 days of data spanning MJD 57500–58500.}

    \label{table1}
\end{table}


\end{document}